\documentclass[12pt]{iopart}
\usepackage{graphicx}

\begin{document}

\title[The effect of RMPs on the divertor in MAST]{The effect of resonant magnetic perturbations on the divertor heat and particle fluxes in MAST}

\author{AJ Thornton$^1$, A Kirk$^1$, P Cahyna$^2$, IT Chapman$^1$, JR Harrison$^1$, Yueqiang Liu$^1$ and the MAST Team}

\address{$^1$ EURATOM/CCFE Fusion Association, Culham Science Centre, Abingdon, Oxon, 0X14 3DB, UK}
\address{$^2$ Institute of Plasma Physics AS CR v.v.i., Association EURATOM/IPP.CR, Prague, Czech Republic}

\ead{andrew.thornton@ccfe.ac.uk}

\begin{abstract}
Edge localised modes (ELMs) are a concern for future devices, such as ITER, due to the large transient heat loads they generate on the divertor surfaces which could limit the operational lifetime of the device. This paper discusses the application of resonant magnetic perturbations (RMPs) as a mechanism for ELM control on MAST. Experiments have been performed using an n=3 toroidal mode number perturbation and measurements of the strike point splitting performed. The measurements have been made using both infrared and visible imaging to measure the heat and particle flux to the divertor. The measured profiles have shown clear splitting in L-mode which compares well with the predication of the splitting location from modelling including the effect of screening. The splitting of the strike point has also been studied as a function of time during the ELM. The splitting varies during the ELM, being the strongest at the time of the peak heat flux and becoming more filamentary at the end of the ELM (200 $\mu s$ after the peak midplane D$\alpha$ emission). Variation in the splitting profiles has also been seen, with some ELMs showing clear splitting and others no splitting. A possible explanation of this effect is proposed, and supported by modelling, which concerns the relative phase between the RMP field and the ELM filament location.

\end{abstract}


\submitto{Nucl. Fusion}

\maketitle

\section{Introduction}
Edge localised modes (ELMs) are a plasma instability driven by steep current and pressure gradients at the plasma edge which lead to the ejection of particles and heat from the plasma. The heat flux to the divertor of ITER as a result of an ELM is projected to be 20 MJ m$^{-2}$ \cite{loarte2007} which could limit the lifetime of the divertor. However, experimental testing of ITER divertor materials under cyclical heat loads \cite{zhitlukin2007} suggest that the threshold for damage is for ELM sizes of  approximately 1 MJ m$^{-2}$ due to the sudden changes in heat flux which compromises the integrity of the material, lowering the tolerance of the material to heat flux as a result of cracking and flaking of the surface. Therefore, there is a requirement for ELM control on ITER to mitigate the heat loads to the divertor surfaces. Several means of ELM control exist, including, vertical kicks \cite{degeling2003}, pellet pacing \cite{lang2004} and the application of external resonant magnetic perturbations (RMPs) \cite{evans2006}. An investigation of the effect of RMPs on the divertor heat loads and particle loads are reported in this paper using data from the Mega Amp Spherical Tokamak (MAST) \cite{lloyd2011}. MAST is equipped with a set of 18 RMP coils, which allow perturbations with a range of toroidal mode numbers to be applied to the plasma. The majority of the results in this paper concentrate on the application of n=3 perturbations to plasmas with double null magnetic configurations.

\begin{figure}[htp]
        \centering
        \includegraphics[width=0.5\textwidth]{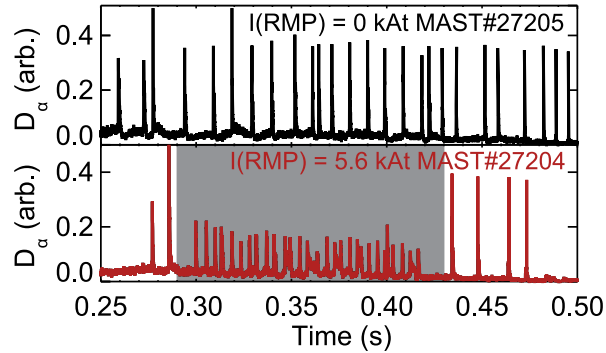}
        \caption{Effect of ELM mitigation via resonant magnetic perturbations (RMPs) on MAST. The top panel shows the midplane D$_{\alpha}$ emission for a discharge without the RMP coils applied and the lower panel shows the effect of applying the RMPs on the ELMs. The coil current is applied in the period marked by the grey shaded window, at the maximum current of 5.6 kAt.}
        \label{fig:elm_mitigation}
\end{figure}

ELM control can take two forms; the ELMs can be completely removed following the application of the RMP in a process known as ELM suppression, or the ELM frequency can be increased which generates smaller ELMs in a process known as ELM mitigation. ELM mitigation has been achieved on MAST, as can be seen in \fref{fig:elm_mitigation}. \Fref{fig:elm_mitigation} shows the midplane D$_{\alpha}$ emission for a RMP off (top panel) and a RMP on (bottom panel) discharge. The ELM frequency can be seen to increase when the RMPs are applied (greyed region on the bottom panel) compared to the RMP off  case with the peak emission also decreasing which suggests the ELMs are smaller. When the RMP coil current is removed from the mitigated discharge, the ELM frequency can be seen to decrease and the emission level returns to that seen prior to the RMPs being applied.

A key issue to address with ELM mitigation experiments is the reduction in divertor heat load seen when the RMP are applied. In addition, observations of the splitting of the strike point offer an indication as to the level of penetration of the RMP field into the plasma and can therefore be used to study the effect of screening due to current flowing along rational surfaces \cite{fitzpatrick1998} or the plasma response to the applied field.

The focus of this paper is an investigation of the effect of the RMP on strike point spatial structure. Section \ref{section:diags} discusses the diagnostics used to measure the heat and particle fluxes to the divertor which are presented in the following sections. The splitting seen in L-mode plasmas are detailed in section \ref{section:strike_pt_split} and compared with vacuum modelling. In H-mode plasmas, the splitting can be measured both inter-ELM and during ELMs, both of which are discussed in section \ref{section:strike_pt_split_hmode}, including the evolution of the splitting as a function of time through the ELM. The response of the plasma to the applied perturbation is studied in section \ref{section:screening}, in terms of plasma screening, and the effect on the modelled strike point splitting. The splitting pattern is seen to vary during the ELM, which is discussed in detail in section \ref{section:variation} with a conclusion to the paper in section \ref{section:conclusion}.

\section{MAST diagnostics}
\label{section:diags}

\begin{figure}[htp]
        \centering
        \includegraphics[width=0.5\textwidth]{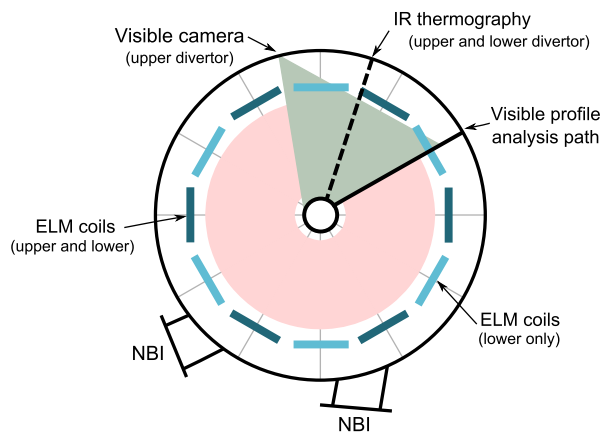}
        \caption{A top down view of the MAST vessel showing the location of the infrared (dashed line) and visible imaging (solid line) analysis paths. The toroidal displacement between the IR analysis path and the visible imaging is approximately 30 degrees.}
        \label{fig:mast_diags}
\end{figure}

The heat flux to the divertor in MAST is routinely measured using an infrared (IR) thermography system. The IR system consists of a medium wave IR (MWIR) camera and a long wave IR (LWIR) camera monitoring the upper and lower divertor surfaces \cite{temmermann2010}. The IR cameras can be operated in two modes; firstly the cameras can be operated at high temporal resolution (up to 15 kHz), and lower spatial resolution (7.5 mm/pixel), in order to measure the heat flux and profile evolution during ELMs. Alternatively, the camera can be operated at a lower temporal resolution, but an increased spatial resolution of 1 mm/pixel for high spatial resolution measurements of the heat flux profile. The location of the IR camera analysis path, relative to the ELM coils and other imaging diagnostics, is shown in \fref{fig:mast_diags} by the dashed line.

In addition to monitoring the divertor heat flux, an estimate of the particle flux to the divertor can be made using filtered visible imaging. Due to the frame rate of the visible imaging being low compared to the repetition rate of the ELMs, estimates of the particle flux can only be made during L-mode and inter-ELM periods. Direct measurements of the particle flux to the divertor using Langmuir probes can be performed, however, the spatial resolution is limited to 1 cm and the temporal resolution only permits analysis of the L-mode and inter-ELM phases. Analysis of the Langmuir probe profiles during L-mode splitting have been reported previously \cite{cahyna2013}. The visible imaging is D$_{\alpha}$ filtered and has a spatial resolution of approximately 1 mm/pixel, similar to that of the high spatial resolution IR imaging. The location of the visible camera, and the field of view is shown in \fref{fig:mast_diags}. The analysis path for the visible imaging is shown by the solid line in \fref{fig:mast_diags} and is displaced from the IR analysis path by approximately 30 degrees toroidally.

\section{Strike point splitting}
\label{section:strike_pt_split}

\begin{figure}[htp]
        \centering
        \includegraphics[width=0.35\textwidth]{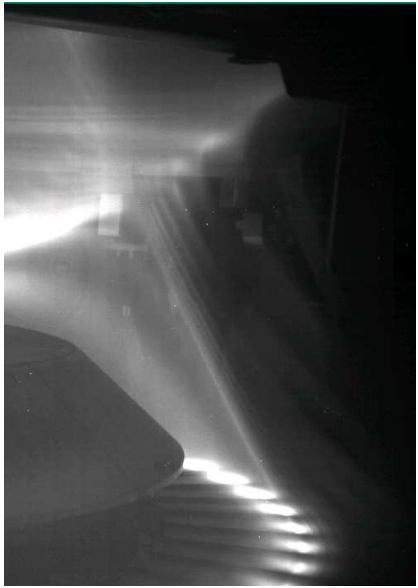}
        \caption{The application of the RMP leads to the formation of X point lobes, these lobes extend down to the divertor leg and strike the divertor surfaces. The figure shows a He II (468nm) filtered image taken from a camera located at the X point in a H-mode plasma, clearly showing the lobes extending to the divertor surface. The interaction of these lobes with the divertor leads to the splitting of the strike point, both in terms of heat flux and particle flux.}
        \label{fig:lobes}
\end{figure}

The application of RMPs breaks the toroidal symmetry of the magnetic field in the tokamak, leading to the production of a three dimensional field. The breaking of the toroidal symmetry of the plasma causes the last closed flux surface (LCFS) to split into two surfaces, known as the stable and unstable manifolds \cite{evans2005}. These surfaces oscillate as they approach the X point and, as a result, form lobe like structures at the outboard and inboard side of the plasma \cite{wingen2009}. The X point lobes can be seen in \fref{fig:lobes} and extend down to the divertor on the outboard side of the plasma (right hand side of the figure) and the lobes on the inboard extend to inner divertor located on the centre column. The interaction of the lobes with the divertor produces splitting of the strike point which has been observed on several machines \cite{jakubowski2009,nardon2011,harting2012}. Measurements of the strike point splitting during the application of the RMP can be made during L-mode and inter-ELM phases in both particle and heat flux. In addition, measurements in the heat flux can be performed during the ELMs. The splitting can then be compared to vacuum modelling predictions of the field line penetration into the plasma.

\subsection{L-mode splitting}

\begin{figure}[htp]
        \centering
        \includegraphics[width=0.5\textwidth]{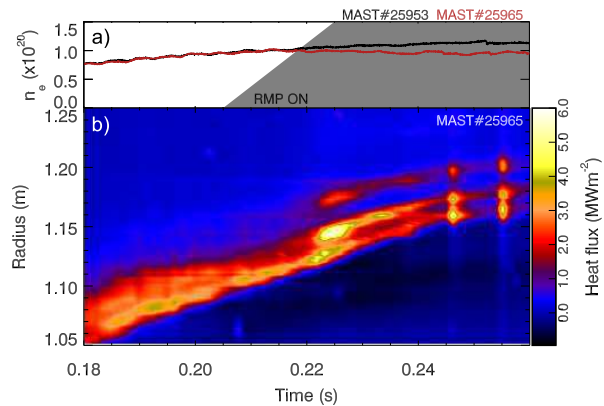}
        \caption{The effect of the RMP on the line integrated density and divertor heat flux in a L-mode plasma. Panel a) shows the line integrated density as a function of time during the discharge, with the greyed region corresponding to the period when the coils are energised. The line integrated density for the RMP off shot is shown in black and the RMP on shot in red. Panel b) shows contour plot of the divertor heat flux as a function of time and radius during a RMP on discharge. The onset of the density pump out and the splitting of the strike point can clearly be seen.}
        \label{fig:scen2_splitting_profile}
\end{figure}

The strike point splitting during L-mode can be measured using the high spatial resolution IR view where the profile resolution is 1 mm. The splitting at the strike point is expected to occur as the RMP coil current is ramped up over a 30 millisecond period, as such, these measurements are made at low temporal resolution of approximately 800 Hz. A variety of discharges have been studied, an example of splitting measurements in a 950 kA double null L-mode plasma is shown in \fref{fig:scen2_splitting_profile}. This discharge has a high divertor heat flux compared to lower plasma current discharges, producing a clear heat flux footprint to the divertor minimising the effect of hot spots on the recorded data \cite{delchambre2009}. 

The temporal and spatial evolution of the divertor strike point is shown in panel b) of \fref{fig:scen2_splitting_profile} as a function of time along the ordinate and radius along the abscissa. The strike point can be seen to sweep across the divertor surface as a function of time during the discharge. The sweeping of the strike point is caused by the solenoid fringing field, which varies during the discharge. The RMP coil current is shown by the shaded region in panel a) along with the line integrated density. As the RMP coil current is increased, the line integrated density can be seen to fall from the level in the RMP off discharge. The drop in density, known as density pump out, is seen at the onset of the strike point splitting and typically characterises the critical threshold for the RMP current in L-mode discharges. Once the RMP current threshold is reached the strike point can be seen to split into three clear lobes which is consistent with the application of a toroidally asymmetric perturbation to the plasma. The formation of the three lobes is also accompanied with a sudden increase in the heat flux to the divertor. The sudden increase in the heat flux to the divertor, at the threshold RMP current value, could be due to the arrival of electrons at the divertor as a result of the formation of a stochastic field at the plasma edge \cite{watkins2009}. 

The modelling of the heat flux pattern is performed using the ERGOS vacuum modelling code \cite{nardon2007_thesis} and assumes no plasma response. The modelling of the divertor footprint involves fieldline tracing from the target to the deepest radius to which they reach, which is quantified in terms of minimum square root normalised flux ($\Psi^{1/2}_{MIN}$). The field line excursion, defined as 1-$\Psi^{1/2}_{MIN}$, can be used as an estimate of the location of the strike point splitting, as regions where there is deep field line penetration should see large heat fluxes to the divertor due to penetration into the core plasma. Regions where the field line excursion is greater than zero correspond to regions where the penetration extends into the confined plasma, and regions where it is less than zero correspond to regions where the field line remains in the scrape off layer (SOL). 

\begin{figure}[ht]
        \centering
        \includegraphics[width=0.7\textwidth]{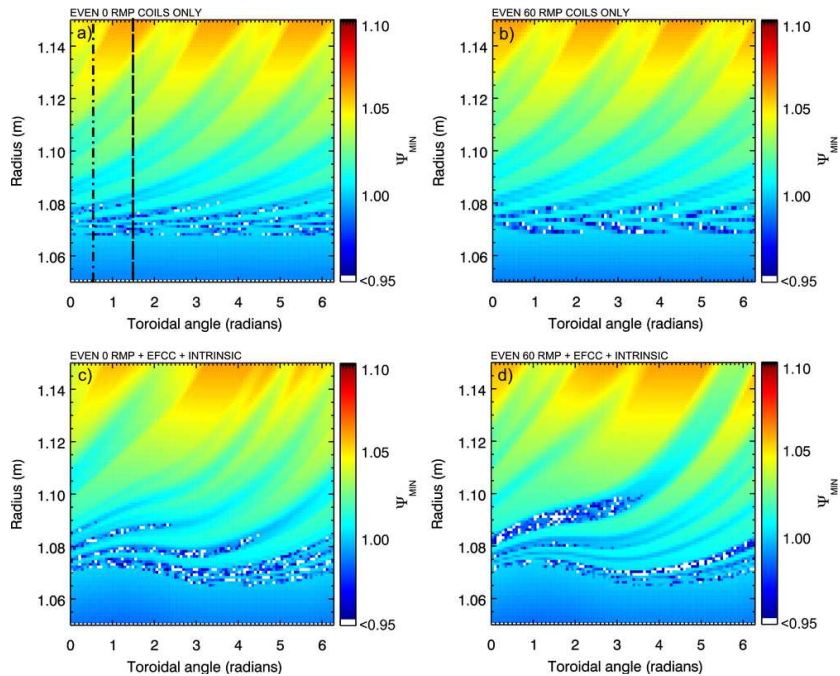}
        \caption{The effect on the modelled strike point pattern at the lower divertor as a function of toroidal angle and radius. Panel a) shows the pattern expected from an even 0 degree phase RMP applied to a 950 kA plasma with panel b) showing the effect of an even 60 degree phase perturbation. The lower panels show the effect of including the error fields present in MAST to the even 0 degree perturbation (panel c)) and the even 60 degree perturbation (panel d)) respectively. The dashed line on panel a) marks the location of the IR profile measurement and the dot-dashed line marks the location of the visible profile measurements.}
        \label{fig:even0_even60_efcc_noefcc}
\end{figure}

The splitting in two coil configurations have been investigated, both with a toroidal mode number of n=3. These coil configurations are an even parity configuration \cite{kirk2010} with a phase of 0 degree or 60 degrees. The phase is relative to the current in the first coil in the machine, which is positive (B$_{r}$ outwards) for the 0 degree phase and negative (B$_{r}$ inwards) for the 60 degree phase. The effect of changing the phase is to rotate the RMP perturbation around the machine, acting to rotate the splitting pattern through the line of sight of the imaging. The effect of changing the phase of the perturbation can be seen by comparing \fref{fig:even0_even60_efcc_noefcc} a) and b) which show the predicted splitting pattern for the two phases. It is also important to include the effect of additional sources of radial field, such as the error field correction coils (EFCCs) and the intrinsic error field into the strike point splitting calculations. The intrinsic error field arises from a misalignment of one of the internal poloidal field coils and has a predominately n=2 component. The magnitude of the intrinsic error field has been measured experimentally using Hall probes during a shutdown period. The field from the EFCC coils is applied externally to the vessel to correct for the intrinsic error field and has an n=2 component from four coils located around the outside of the MAST vacuum vessel.The inclusion of these additional fields have a significant effect on the strike point pattern as can be seen in  \fref{fig:even0_even60_efcc_noefcc} c) and d) which shows the modelled strike point pattern for the even 0 degree and 60 degree cases respectively.

\begin{figure}[htp]
    \centering
    \includegraphics[width=0.5\textwidth]{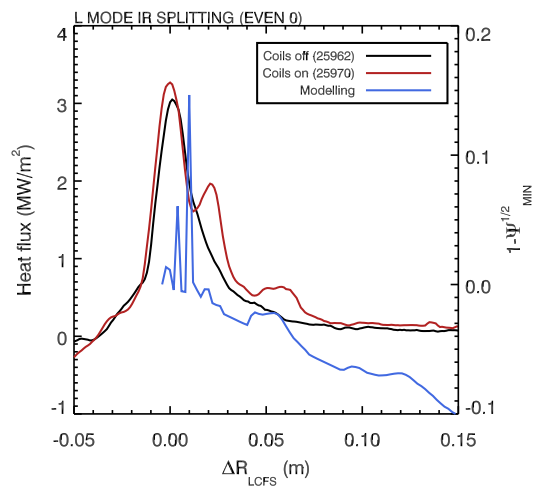}
    \caption{Comparison of the infrared splitting at the lower divertor in the case of a 950kA L-mode discharge. The applied perturbation has a toroidal mode number of n=3 and a phase of 0 degrees. The data is shown for both RMP on and RMP off and compared with vacuum modelling.}
    \label{fig:scen2_even0}
\end{figure}

\begin{figure}[htp]
	\centering
    \includegraphics[width=0.5\textwidth]{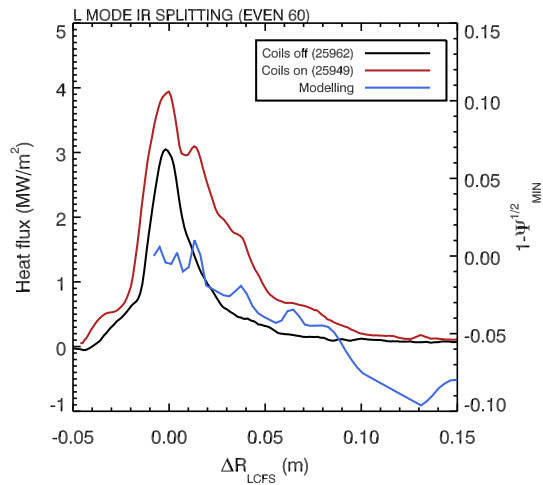}
    \caption{The infrared splitting measured in a 950kA L-mode discharge with an n=3, 60 degree phase coil configuration. Profiles of the RMP on and RMP off splitting are shown, along with modelling of the expected pattern from vacuum modelling.}
    \label{fig:scen2_even60}
\end{figure}

Comparison of the two phases of the applied perturbation can be made using profiles extracted from each of the two phases at the correct toroidal angle corresponding to the IR and visible cameras. These profiles, taken from \Fref{fig:even0_even60_efcc_noefcc} c) and d), are shown in \fref{fig:scen2_even0} and \fref{fig:scen2_even60} alongside the measured profiles. \Fref{fig:scen2_even0} shows the splitting in the case of the even 0 configuration, with the IR profile being clearly split into three lobes, as expected from the applied toroidal mode number of the perturbation. The location of the splitting can be compared with vacuum modelling predictions of the field line excursion which is plotted against the right hand axes in the figures. The outer lobes in \fref{fig:scen2_even0} at $\Delta R_{LCFS} = 0.02$ and 0.05 m are well matched with the measured profile. The modelling shows three lobes with large field line penetration at $ 0 < \Delta R_{LCFS} < 0.02 m$ could correspond to the heat flux at the LCFS considering that the vacuum modelling does not include the effect of cross field diffusion, or instrument function of the camera which will act to blur the closely spaced lobes. The vacuum calculations do not include the effect of the plasma response. The inclusion of the plasma response will be discussed in section \ref{section:screening} and can explain the large lobe at $\Delta R_{LCFS} = 0.01$ m which is not seen in the IR profiles. In contrast, the splitting in the IR profile for the even 60 case is smaller, as shown in \fref{fig:scen2_even60}. The smaller splitting in the even 60 case is caused by two factors; firstly the width of the splitting varies as a function of toroidal angle and secondly the interaction between error field and the RMP will change depending on the relative phases of the RMP and error field. The reduced splitting is measured by the IR camera and supported by the field line excursion calculations, as the field line excursion for the lobes in the even 60 case being smaller in magnitude than that in the even 0 case. In the even 60 case, the location of the lobes and the deepest regions of field line excursion are well matched for the lobes with $\Delta R_{LCFS} < 0.1$ m which is also the case in the even 0 case. However, the lobe predicted at $\Delta R_{LCFS} = 0.12 $ m in the even 0 case and $\Delta R_{LCFS} = 0.15$ m in the even 60 case is not seen in the IR profiles. The location of these lobes at the divertor correspond to a region approximately 3 cm outside the LCFS at the midplane. Thomson scattering measurements of the temperature and density at the midplane show 1 cm fall off lengths in both of these profiles, it is expected that little plasma will reach these lobes explaining the absence of the outer most lobes in the L mode profiles.

\begin{figure}[htp]
        \centering
        \includegraphics[width=0.5\textwidth]{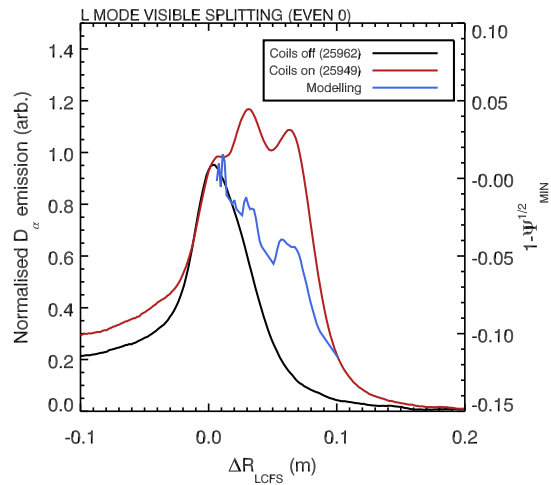}
        \caption{Splitting of the strike point in particle flux, as measured by filtered visible imaging in the n=3, 0 degree phase configuration. The profiles are shown for a RMP on, RMP off and modelled profile.}
        \label{fig:scen2_vis_even0}
\end{figure}

\begin{figure}[htp]
        \centering
        \includegraphics[width=0.5\textwidth]{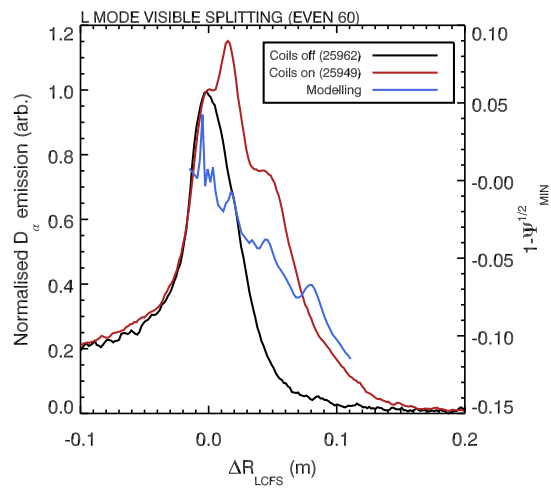}
        \caption{Splitting of the strike point in particle flux, as measured by filtered visible imaging in the n=3, 60 degree phase configuration. The profiles are shown for a RMP on, RMP off and modelled profile.}
        \label{fig:scen2_vis_even60}
\end{figure}

Visible imaging of the divertor allows measurements of the particle flux profiles to the divertor to be made in these two configurations. The visible profiles for the even 0 configuration are shown in \fref{fig:scen2_vis_even0} and for the even 60 configuration in \fref{fig:scen2_vis_even60} and confirm the results seen in the IR data. The visible profiles have been normalised in magnitude to the peak at the separatrix to allow the RMP on and RMP off profiles to be compared. The location of the lobes in the visible data is expected to differ from the IR data as the two cameras measure at different toroidal locations. There is clear splitting of the particle flux into three lobes, as observed in the IR profiles. However, the largest emission is seen on the secondary lobe of the splitting in the visible imaging, this has also been seen in particle flux profiles on other devices \cite{schmitz2008}. The visible emission will be determined by the recombination of the plasma in the strike point region, therefore the higher secondary emission could result from a more optimal plasma temperature in this region for recombination. The comparison of the modelled strike point field excursion is in moderate agreement with the measured profiles.The lower level of splitting seen in the even 60 case compared to the even 0 case is also reflected in the visible imaging data.

\section{H-mode splitting}
\label{section:strike_pt_split_hmode}

The splitting in H-mode can be separated into two phases; the inter-ELM phase, where the splitting is generated only by the applied perturbation, and during ELMs when the splitting from the applied perturbation is affected by the filamentary nature of the ELMs. The H mode measurements are performed in a 600 kA double null discharge which exhibits regular type I ELMs without the application of RMP. Inter-ELM filaments \cite{ayed2009} are not detected in the visible and IR measurements presented here. In the case of the visible imaging, the low temporal resolution prevents the inter-ELM filaments from being resolved as this requires frame rates in excess of 10 kHz. The IR camera has a frame rate compatible with imaging the inter-ELM filaments, however, profiles are averaged and the heat flux contained within the inter-ELM filament is too small to give rise to significant divertor heating above the static inter-ELM heat flux.

\begin{figure}[htp]
        \centering
        \includegraphics[width=0.5\textwidth]{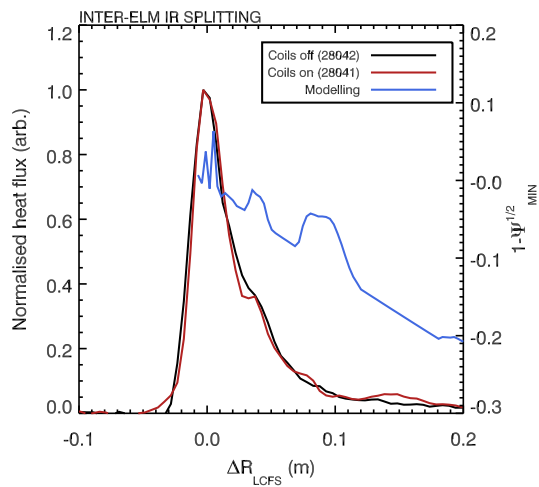}
        \caption{Infra-red profiles of the divertor heat flux for times both before and during the application of the RMP coils. The modelled profile (blue) shows the field line excursion calculated using vacuum modelling.}
        \label{fig:scen6_inter_elm}
\end{figure}

\begin{figure}[htp]
        \centering
        \includegraphics[width=0.5\textwidth]{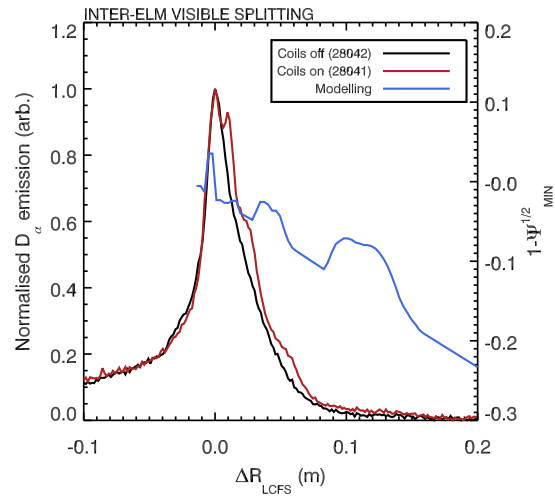}
        \caption{Visible imaging profiles of the divertor particle flux for a time both before and during the application of the RMP coils. The modelled profile is shown as plotted along the right hand axis.}
        \label{fig:scen6_vis_inter_elm}
\end{figure}

Considering the inter-ELM case first, the visible and IR profiles can be extracted from the periods between ELMs and averaged together to study the splitting due to the RMP coils. The profiles for the inter-ELM IR splitting are shown in \fref{fig:scen6_inter_elm} and the visible profiles in \fref{fig:scen6_vis_inter_elm} for the application of an n=3, even 0 degree phase perturbation. The configuration was chosen as it produced the largest splitting in the L-mode case. As was the case for the L-mode profiles, the measured profiles are shown alongside the modelled field line excursion.

\begin{figure}[ht]
        \centering
        \includegraphics[width=0.4\textwidth]{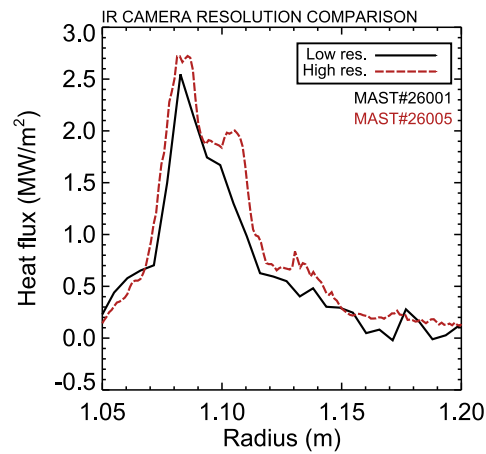}
        \caption{Comparison of the two spatial resolutions used to collect IR profiles in RMP on discharges. The data is taken from two repeated L-mode discharges, in which the IR profile has been measured using both the low (black solid) and high resolution (red dashed) camera views at the same time in the discharge and using the same RMP coil current.}
        \label{fig:camera_resolutions}
\end{figure}

\Fref{fig:scen6_inter_elm} shows the IR profiles from an RMP off and RMP on inter-ELM H mode case. The small shoulder in the RMP off case at $\Delta R_{LCFS}$ = 0.04 is due to a tile gap in the profile. Upon application of the RMP, a clear lobe is seen at $\Delta R_{LCFS}$ = 0.04 m and a lobes at $\Delta R_{LCFS}$ = 0.08m is formed which is not present in the RMP off case. It is clear from the figure that the inter-ELM IR splitting is less clearly defined than in the L-mode case. It should be noted that the spatial resolution of the inter-ELM data is lower compared to that of the L-mode data, as a result of the increased temporal resolution required to measure the inter-ELM periods and separate them from the ELMs. The effect of the decreased spatial resolution can be seen in \fref{fig:camera_resolutions}, which shows the splitting in an L-mode discharge measured using both the high and low spatial resolutions. 

The outer most lobe in the vacuum modelled field line excursion is not visible in the experimental profiles and a lobes is present in the RMP on profile at $\Delta R_{LCFS}$ = 0.15 m is seen to form which is not in the modelled profile. The magnitude of the lobe at $\Delta R_{LCFS}$ = 0.15 m is around half that of the lobes at $\Delta R_{LCFS}$ = 0.08 m. Moderate agreement is seen between the measured IR profile and the modelled field line excursion, with the exception of the region beyond $\Delta R_{LCFS}$ = 0.12 m, and further measurements at increased spatial resolution and lower noise levels would be required to determine the origin of the lobe at $\Delta R_{LCFS}$ = 0.15 m.

The improved spatial resolution of the visible imaging produces profiles with increased detail of the structure compared to the IR measurements. \Fref{fig:scen6_vis_inter_elm} shows the visible profiles measured during the inter-ELM period in the RMP off and RMP on cases. The field line excursion for vacuum modelling is also shown on the right hand axis. There is moderate agreement between the modelled profile and the measured splitting in the visible case, the outermost lobe in the modelled profile is not present in the measured profile as in the case of the IR profiles, which will be discussed in section \ref{section:screening}.

\begin{figure}
		\centering
        \includegraphics[width=0.5\textwidth]{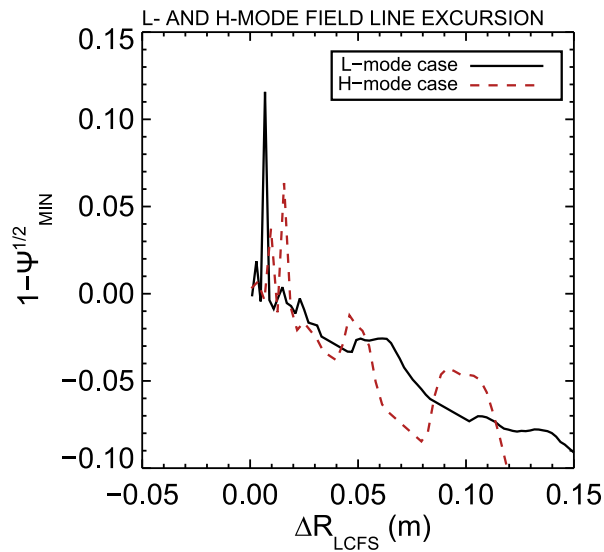}
        \caption{Field line excursion profiles calculated for the L-mode (black, solid line) and H-mode (red, dashed line) using vacuum modelling.}
        \label{fig:lh_field_excursion}
\end{figure}

It is clear that there is a large difference between the splitting seen in the L-mode case and the H-mode case. The data for these two cases are taken from discharges with different plasma scenarios and, as result, the penetration of the fields is different in each case. The effect of the different plasma scenarios on the field line excursion can be seen in \fref{fig:lh_field_excursion} which shows the modelled field line excursion for each of the plasmas, as a function of distance from the LCFS. The H-mode case can be seen to have splitting over a smaller range, approximately 10 cm, compared to the L-mode case where splitting is predicted over a 15 cm range. In addition, there is lower penetration of the field lines into the confined region of the plasma in the H mode case and more field lines are confined to the SOL region compared to the L mode case which will act to narrow the heat flux pattern in the H mode case. The decreased field line penetration into the confined region, along with the decreased cross field transport present in H mode could both act to decrease the heat flux in the outer lobes of the profiles and explain the decreased level of splitting measured in the H mode case compared to L mode. A full model of the divertor heat flux, including transport and field line effects would be required to confirm this hypothesis.

\section{Effect of the plasma response on the divertor profiles}
\label{section:screening}

\begin{figure}[htp]
        \centering
        \includegraphics[width=0.6\textwidth]{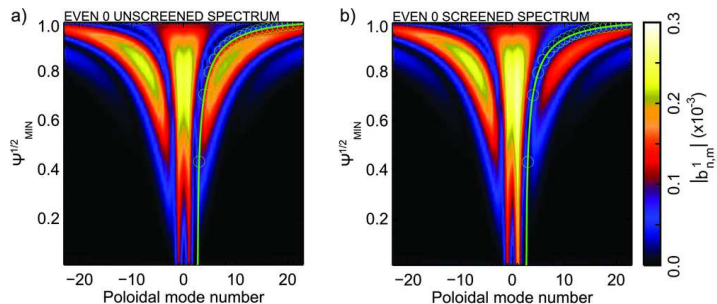}
        \caption{Poloidal magnetic spectra of the even 0 configuration applied to a 950kA discharge for a) the vacuum field case and b) the case with plasma screening at the rational surfaces using an ideal MHD response. The green line is defined by the safety factor multiplied by the toroidal mode number of the applied perturbation (n=3).}
        \label{fig:spectrum_screen}
\end{figure}

The plasma response can act to either screen out the applied RMP field or enhance it. One form of plasma response, known as rotation screening \cite{fitzpatrick1998},  originates from the plasma rotating through the static RMP field. The rotation of the plasma through the RMP field is thought to induce currents to flow which act to screen the applied perturbation from the plasma. The effect of the plasma screening can be included using an ad-hoc screening code \cite{cahyna2011} which introduces helical currents on the rational surfaces to cancel the radial magnetic field generated by the RMP. The effect of the screening can be seen by comparing poloidal magnetic spectra in \fref{fig:spectrum_screen} a) (vacuum case) and \fref{fig:spectrum_screen} b) (screened case) which show the normalised perpendicular component of the perturbed field, $|b^{1}_{mn}|$, as a function of poloidal flux and mode number, $m$. The normalised component of the perturbed field is defined as $|b^{1}_{mn}| = \left ( \vec{B} \cdot \vec{\nabla} \Psi_{pol}^{1/2} \right ) / \left ( \vec{B} \cdot \vec{\nabla} \phi \right )$, where $\vec{B}$ is the total B field vector, $\Psi_{pol}^{1/2})$ is the poloidal flux and $\phi$ is the toroidal angle \cite{nardon2007_thesis}.

\Fref{fig:spectrum_screen} a) shows the location of the rational surfaces in the plasma (green circles), and these points can be seen to coincide with the upper edge of the resonance in $|b^{1}_{mn}|$. The screening currents are applied to cancel the normalised perpendicular component of the perturbed field at the rational surfaces. The effect of applying the screening to surfaces with n=3 and $3 \leq m \leq 16$ is shown in \fref{fig:spectrum_screen} b). The reduction in the $|b^{1}_{mn}|$ can be seen along the location of the rational surfaces in the plot. A valley is formed in the spectrum, where the perturbed field component is reduced by the currents applied on the rational surfaces. 

\begin{figure}[htp]
        \centering
        \includegraphics[width=0.5\textwidth]{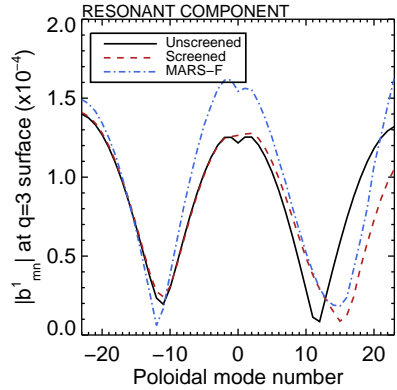}
        \caption{The normalised perpendicular component of the perturbed field, $|b^{1}_{mn}|$, on the q$_{95}$ surface, for the unscreened (black solid), screened (red dashed) and MARS-F (blue dot dashed) n=3 even 0 RMP fields in a 950 kA discharge. The screened field can be seen to reduce the field component at the m=14 surface compared to the unscreened case.}
        \label{fig:b1mn_qsurf}
\end{figure}

The ad-hoc screening model assumes ideal screening of the RMP field from the plasma and does not include the effect of plasma rotation or amplification of the RMP field by the plasma. MARS-F \cite{liu2010} is a linear single fluid resistive MHD code which models the plasma response to the RMP field via the effect of screening from toroidal rotation. A comparison can be made between the vacuum, ad-hoc screened and MARS-F plasma response models. The effect of the plasma response can be clearly seen in \fref{fig:b1mn_qsurf} which shows the $|b^{1}_{mn}|$ component in the region of the q$_{95}$ surface for the screened, MARS-F and unscreened cases. The $|b^{1}_{mn}|$ component of the applied perturbation (n=3) can be seen to be significantly reduced in the region around the m=14 surface in the plasma response cases (screened and MARS-F). The reduction effectively reduces the component to zero at the m=14 surface within the numerical resolution used in the modelling. The effect on the adjacent surfaces can also be seen, as the magnitude of the $|b^{1}_{mn}|$ component is reduced on surfaces where m $>$ 14 for the screening model. The effect of plasma amplification of the field can be seen on adjacent surfaces ($-10 < m < 14$) in the MARS-F case.

\begin{figure}[htp]
        \centering
        \includegraphics[width=0.5\textwidth]{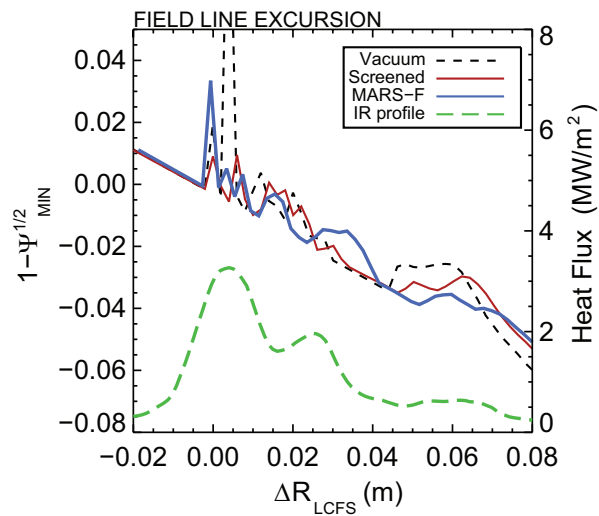}
        \caption{Modelled strike point splitting pattern for a vacuum case (black, short dashed line), ad-hoc screened plasma response (red, short dashed line) and MARS-F plasma response (blue, dot dashed line) showing the effect of the plasma response on the strike point profiles. The measured IR profile is also shown for comparison with the modelled data (green, long dashed line)}
        \label{fig:screened_footprint}
\end{figure}

The reduction in the applied field due to the plasma response will affect the divertor strike point pattern compared to the vacuum field calculations shown in the previous sections. It has been shown that the screening reduces the lobe length in the toroidal direction \cite{cahyna2011}, this decreases the number of lobes seen in a radial profile at a single toroidal angle. The result of modelling the strike point pattern as a function of toroidal angle in the cases shown in this paper supports this observation. A single profile of the field excursion, at the toroidal angle of the IR camera, can be extracted to allow a direct comparison of vacuum and screened profiles from an ad-hoc screening model and from MARS-F simulation. The modelled strike point splitting pattern in the L-mode 950 kA discharge shown in \fref{fig:scen2_even0} using a vacuum model, is now shown in \fref{fig:screened_footprint} including modelling with and without the effect of plasma screening. The figure shows a clear reduction in the depth of penetration for the field lines which originate in the core ($1-\Psi_{MIN}^{1/2} > 0$), comparing the vacuum case to the screened case there is a reduction in the field line excursion at the location of the largest lobe ($\Delta R_{LCFS} = 0.01 m$) of 95\%. Those lobes which are confined to the SOL region, ($1-\Psi_{MIN}^{1/2} < 0$), show some changes in the overall shape and a reduction in magnitude of the lobe, but the lobes are still present in the field line excursion in both the ad-hoc screening model or MARS-F model. The SOL lobes are localised to the region outside the plasma through the interaction of the RMP field and the vacuum field from the plasma. This is a region where the field lines are open and, as a result, no screening currents can form.

The plasma response can be seen to affect the lobes where the field line excursion is greater than one, and acts to screen them out of the plasma. The vacuum modelling shown in \Fref{fig:scen2_even0} shows a large lobes should be present at $\Delta R_{LCFS} = 0.01$ m which extends to a field line excursion above 0.1 in the vacuum case. There is no large lobe measured in the IR at this location, although it could be the case that the instrument function of the IR camera merges this lobe with the adjacent ones. The plasma response modelling of the splitting in \fref{fig:scen2_even0} is shown in \fref{fig:screened_footprint}. It is clear that the amplitude of the lobe at $\Delta R_{LCFS} = 0.01 m$ is significantly reduced by the effect of the plasma response by comparing the short dashed line for vacuum modelling to the solid lines for the model including the plasma response. This suggests that the absence of this lobe in the IR profiles may not be due to the IR spatial resolution alone, but due to the effect of the plasma response on the penetration of the applied RMP field into the plasma. The SOL lobes are largely unaffected by the screening of the applied field, with small reductions in the field line excursion seen on these lobes from the MARS-F modelling. In order for the SOL lobes to be visible, there must be sufficient cross field transport to deposit particles and heat into the lobes upstream of the divertor. 

\section{Splitting during ELMs}
\label{section:elm_split}

\subsection{Measurements during ELMs with and without RMPs}

\begin{figure}[htp]
        \centering
        \includegraphics[width=0.5\textwidth]{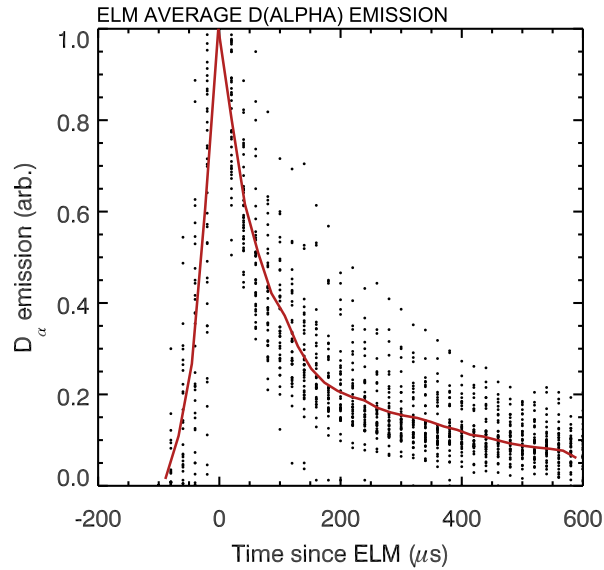}
        \caption{Coherent average of the midplane D$_{\alpha}$ emission from the ELMs. The peak of the D$_{\alpha}$ emission is taken as the start of the ELM.}
        \label{fig:avg_dalpha}
\end{figure}

\begin{figure}[htp]
        \centering
        \includegraphics[width=0.5\textwidth]{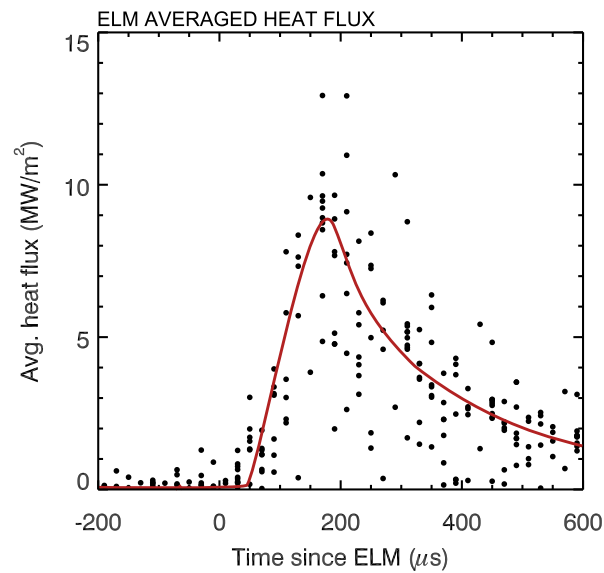}
        \caption{The ELM heat flux coherently averaged for several ELMs. The time base is taken relative to the peak of the midplane D$_{\alpha}$ emission.}
        \label{fig:avg_qpeak}
\end{figure}

ELM filaments first form at or near the LCFS and remain there for the first 50 to 100$\mu$s after the start of the ELM \cite{kirk2009}. The filaments rotate with the bulk plasma during this phase, allowing energy and particles to leave the core plasma and be deposited at the divertor. The connected filament increases the width and magnitude of the heat flux arriving at the divertor \cite{kirk2007}. At 100 $\mu$s after the start of the ELM, the filament separates from the plasma edge and accelerates away radially, whilst decelerating toroidally. During this phase, the energy and particles contained within the filament are deposited onto the divertor as the filament propagates radially outwards \cite{kirk2005}. The loss of energy from the filament will be controlled by the parallel transit time and the radial motion of the filament, as a result the heat flux at the divertor will vary as a function of time during the emission of the ELM. The evolution of the heat flux to the divertor can be followed as a function of time during the ELM. The temporal evolution of the D$_{\alpha}$ emission from an ELM is shown in \fref{fig:avg_dalpha}, where the profile is generated by coherently averaging over several ELMs. The ELM time is defined as the peak D$_{\alpha}$ emission at the midplane, with all times being relative to this peak. The delay between the ELM occurring at the midplane and the heat flux arriving at the divertor can be seen in \fref{fig:avg_qpeak} which shows a coherent average of the ELM heat flux. The peak of the heat flux occurs 150 $\mu s$ after the peak of the D$_{\alpha}$. The observed delay is consistent with the parallel ion transit time in MAST for propagation from the midplane to the divertor.

\begin{figure}[htp]
        \centering
        \includegraphics[width=0.7\textwidth]{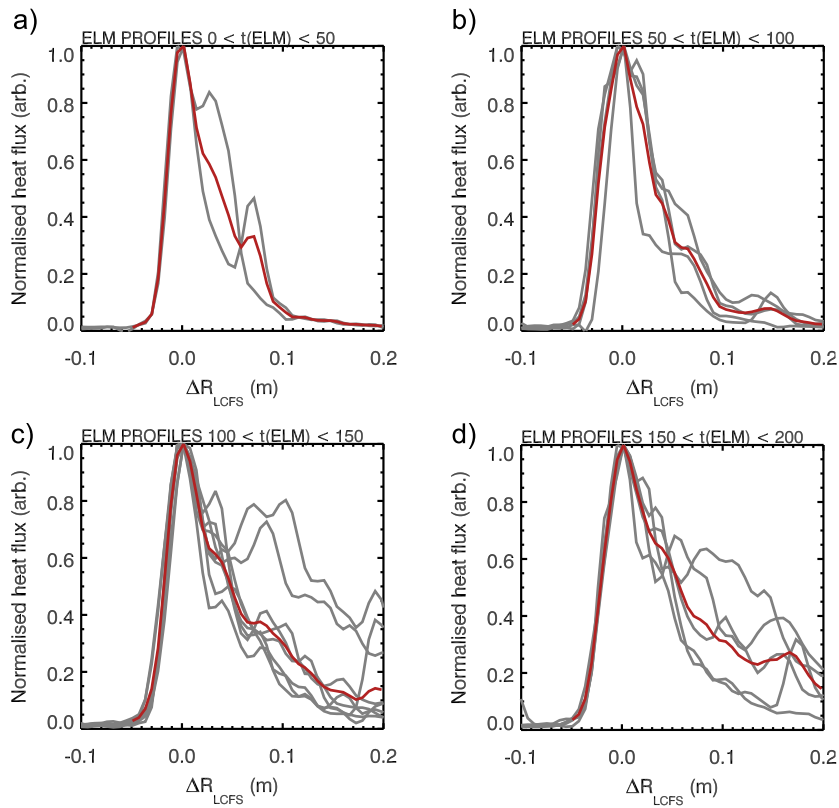}
        \caption{Infra red profiles from ELMs grouped as a function of time through the ELM for unmitigated (RMP off) ELMs. The profiles are averaged in groups of 50 $\mu s$. The grey profiles are from individual ELMs and the red profiles are averages of the grey profiles.}
        \label{fig:elm_cycle_profiles_off}
\end{figure}

The particles and energy are lost from the ELM filament as it propagates outward from the LCFS. The motion of the filament across the magnetic field causes the location where the energy is deposited onto the divertor to move in time. The deposition from a single ELM exhibits a spiral pattern and shows many striations, which resemble splitting of the strike point \cite{kirk2007}. As the toroidal location at which an ELM is ejected varies from ELM to ELM, the splitting seen in natural, unmitigated ELMs will coherently average to a smooth profile due to the filamentary nature of the ELMs \cite{kirk2006}. The IR profile from a series of coherently averaged natural ELMs can be seen in \fref{fig:elm_cycle_profiles_off} where the profiles from individual ELMs are shown in grey and an average of all of the profiles in red. It can be seen that the grey profiles show striations at locations which vary from ELM to ELM, resulting in no coherent structures being observed in the averaged profile.

\begin{figure}[htp]
        \centering
        \includegraphics[width=0.7\textwidth]{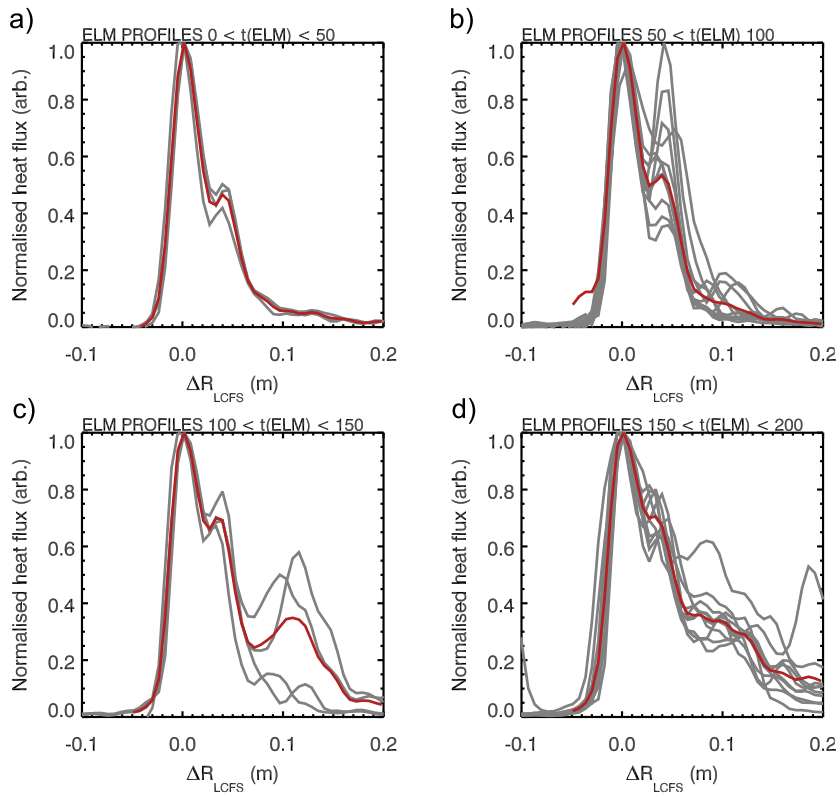}
        \caption{Infra red profiles from ELMs grouped as a function of time through the ELM for mitigated (RMP on) ELMs. The profiles are averaged in groups of 50 $\mu s$. The grey profiles are from individual ELMs and the red profiles are averages of the grey profiles. The inter-ELM data shows splitting of the strike point at $\Delta R_{LCFS}$ = 0.04 and 0.08m, which is consistent with the splitting seen in the profiles during the ELMs.}
        \label{fig:elm_cycle_profiles}
\end{figure}

The application of the RMP changes the IR profiles obtained during the ELM when compared to the RMP off case. \Fref{fig:elm_cycle_profiles} shows the coherently averaged profiles in the RMP on case, averaged over 50 $\mu s$ windows during the ELM. The averaged profiles in \fref{fig:elm_cycle_profiles} show clear splitting compared to the profiles in \fref{fig:elm_cycle_profiles_off} in which the splitting seen in individual ELMs averages away. At the start of the ELM, as shown in \fref{fig:elm_cycle_profiles} a), there is clear splitting in the profile, with the formation of a secondary lobe in the SOL region of the profile. It should be noted that there is a tile gap at this location, however, the RMP on profiles show a clear, consistent enhanced heat flux at this location compared to the RMP off profiles, which suggests that a lobe also forms at this location. Also, this lobe is not consistently present in the RMP off case (\fref{fig:elm_cycle_profiles_off} a)). As the ELM evolves, the magnitude and number of lobes vary, as shown in \fref{fig:elm_cycle_profiles} b) where the secondary lobe strengthens and there is evidence for the formation of a third lobe at $\Delta R _{LCFS}$ = 0.08m. The profiles shown in \fref{fig:elm_cycle_profiles} c) are taken at the time of the peak ELM heat flux at the divertor, where three lobes can clearly be seen in the profiles. Investigation of the splitting on DIII-D \cite{jakubowski2009} has shown that as the ELM energy loss increases, an increasing number of lobes are seen at the divertor. The increased ELM energy delivers a larger heat flux to the divertor, which deposits more energy into the lobes that extend to the divertor giving splitting. A similar result is seen in MAST, whereby the increasing heat flux through the ELM acts to deposit more energy into the outer lobes of the splitting making them visible at the divertor. 

\begin{figure}[ht]
        \centering
        \includegraphics[width=0.7\textwidth]{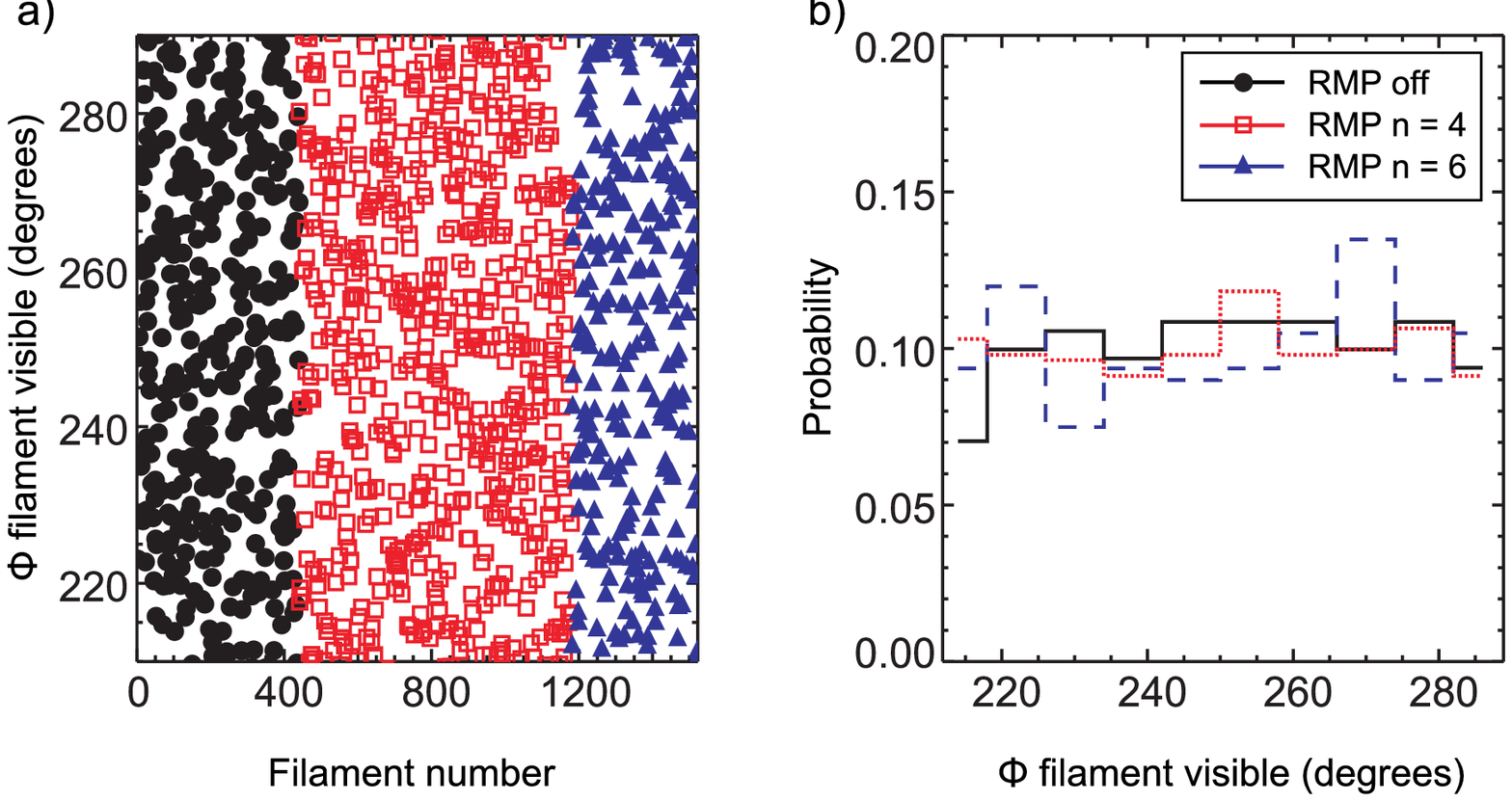}
        \caption{The toroidal angle at which the ELM filaments are first visible for a range of RMP configurations. Panel a) shows the toroidal angle at which filaments are born for a range of different RMP configurations; RMP off (black circles), n=4 RMP (red squares) and n=6 RMP (blue triangles). Panel b) shows a histogram of the toroidal angle for each of the configurations, it can be seen that there is no preferred toroidal angle at which the filaments originate.}
        \label{fig:filament_start}
\end{figure}

The final set of averaged profiles in \fref{fig:elm_cycle_profiles} d) start to show a loss of coherent structure at the strike point. The loss of the splitting at this time could result from the arrival of the ELM filaments at the divertor which act to spread out and randomise the location of the striations in the heat flux at the divertor. In order for this to be the case, the filaments must be emitted at random locations when the RMP are applied and are therefore not locked to the applied RMP field. Previous studies \cite{kirk2013_2} using a range of RMP mode numbers have collected fast imaging data of the filaments. The fast imaging of the filaments can be used to determine if the filaments are emitted at random locations, by tracking the toroidal angle at which they are emitted. The toroidal angle of the filaments is determined by registering the camera view with the vessel, enabling the pixel number on the camera to be equated to a given angle. The angle at which the filaments are first visible can then be plotted for a number of filaments, as shown in \fref{fig:filament_start} a). \Fref{fig:filament_start} a) shows that in the RMP off case and RMP on cases (for an n=4 and n=6 RMP toroidal mode number) the starting location of the filament is random in the field of view of the camera. A histogram of the toroidal angle of the filaments, \fref{fig:filament_start} b), shows that there is an even distribution of the filaments across all of the toroidal angles in the field of view of the fast camera. These results confirm that the filaments are not locked to the applied RMP and support the IR data which shows the random arrival of filaments at the divertor at the end of the ELM event.

\begin{figure}[htp]
        \centering
        \includegraphics[width=0.5\textwidth]{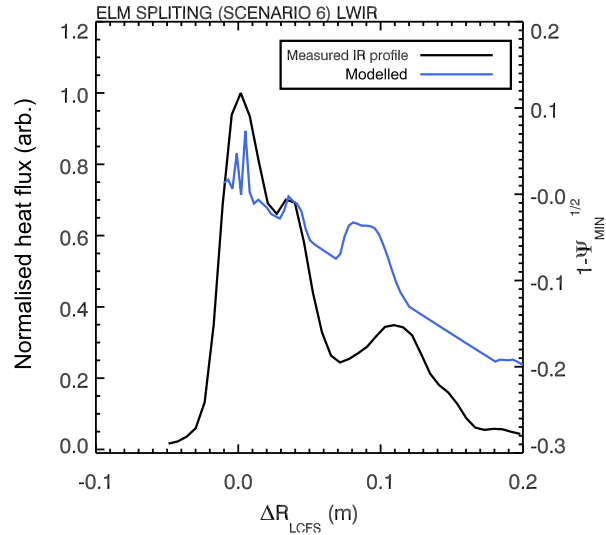}
        \caption{The averaged infra red profile at the peak of the ELM heat flux compared to vacuum modelling of the splitting. There is some variation in the q$_{95}$ value during the period over which the ELM profiles are averaged.}
        \label{fig:elm_modelled_cf}
\end{figure}

The measured ELM profile at the time of the peak ELM heat flux can be compared to vacuum modelling, as shown in \fref{fig:elm_modelled_cf}. There is good agreement between the lobes in the modelled profile and the measured profiles, some mismatch is seen in the location of the outermost lobe. The q$_{95}$ value changes as a function of time through the discharge, and the ELM profiles are averaged during this period. The variation in q$_{95}$ will cause the splitting pattern to vary, which could be a cause of the poor match between the modelling and the measured outer lobe. It should also be noted that the splitting in the case of the ELMs shown in \fref{fig:elm_cycle_profiles} and \fref{fig:elm_modelled_cf} occurs at a location that is similar to the measured splitting in the inter-ELM case, as shown in \fref{fig:scen6_inter_elm}, where the splitting is seen at $\Delta R_{LCFS}$ = 0.04 m and 0.08 m in comparison to $\Delta R_{LCFS}$ = 0.04 m and 0.1 m for the ELM splitting case.

\subsection{Variability in ELM splitting}
\label{section:variation}

\begin{figure}[htp]
        \centering
        \includegraphics[width=0.4\textwidth]{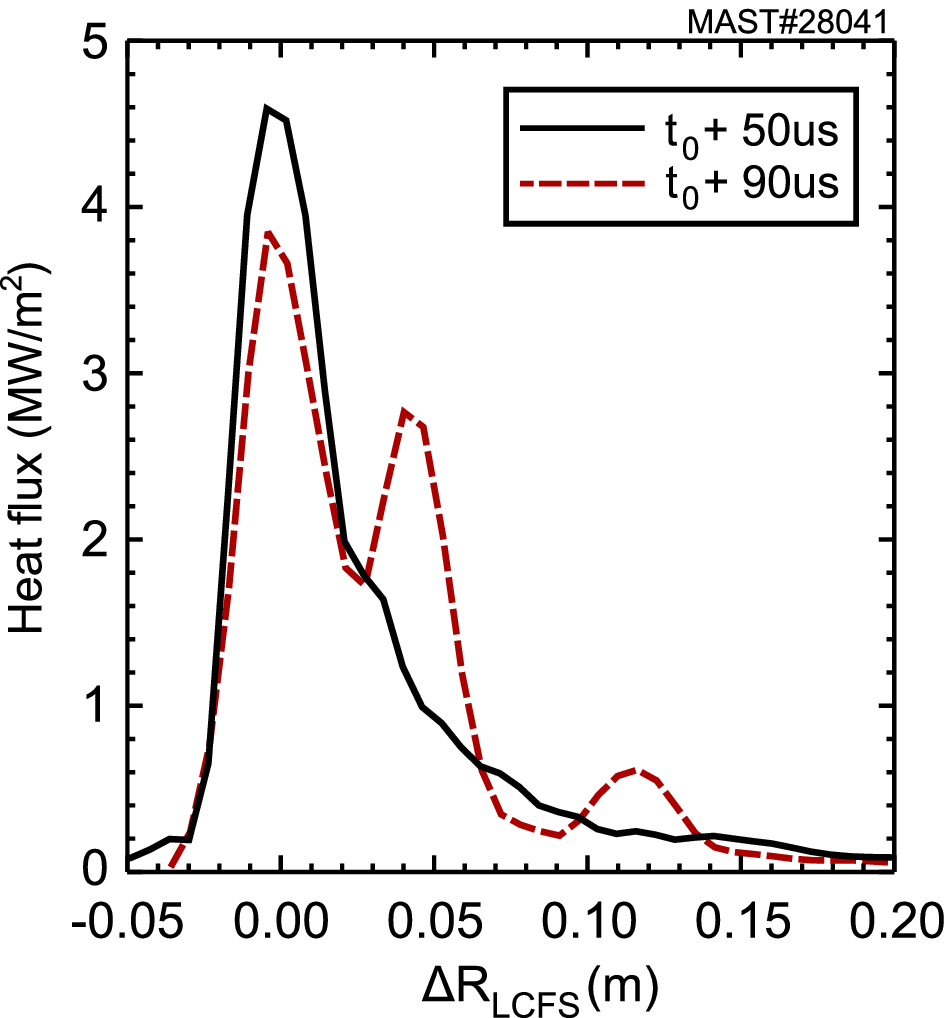}
        \caption{Infra red profiles taken at similar periods during the ELM (50 to 100 us after the peak of the D$\alpha$ emission) of an RMP applied discharge, one showing no splitting (black solid line) and one showing clear splitting (dashed red line).}
        \label{fig:nosplit_split_profiles}
\end{figure}

The ELM profiles in RMP on discharges show variability in the level of splitting from ELM to ELM, even at the same point during the ELM. In some cases early in the ELM splitting is always seen at the same place, or no splitting is seen at all. \Fref{fig:nosplit_split_profiles} shows two examples of profiles taken at similar times during the ELM, one of which shows clear splitting and one which show no evidence for splitting.  

\begin{figure}[htp]
        \centering
        \includegraphics[width=0.35\textwidth]{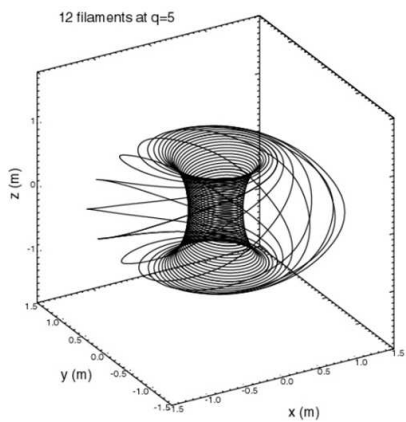}
        \caption{Representation of the ELM filaments for the modelling. The location of the filaments in space is set be a field line located at the q$_{95}$ surface.}
        \label{fig:n12_elm_filaments}
\end{figure}

The profiles which exhibit clear splitting are selected for analysis in the previous section, typically this is around one third of the profiles collected. One possible cause for the variation is the alignment between the ELM peeling ballooning mode and the phase of the RMP. It has been shown that the ELM is composed of approximately 12-16 filaments \cite{kirk2011}, each of which can carry a current to the divertor \cite{alladio2008, evans2009}. As the filaments carry a current of between 200 and 300A, \cite{kirk2006} and are toroidally localised, they act as a perturbing field with a toroidal mode number, $n$, of 12-16. The effect of the ELM filaments on the strike point splitting pattern can be modelled by introducing 12 current carrying filaments aligned with the magnetic field at the q$_{95}$ surface, as shown in \fref{fig:n12_elm_filaments}. 

\begin{figure}[htp]
        \centering
        \includegraphics[width=0.45\textwidth, angle=90]{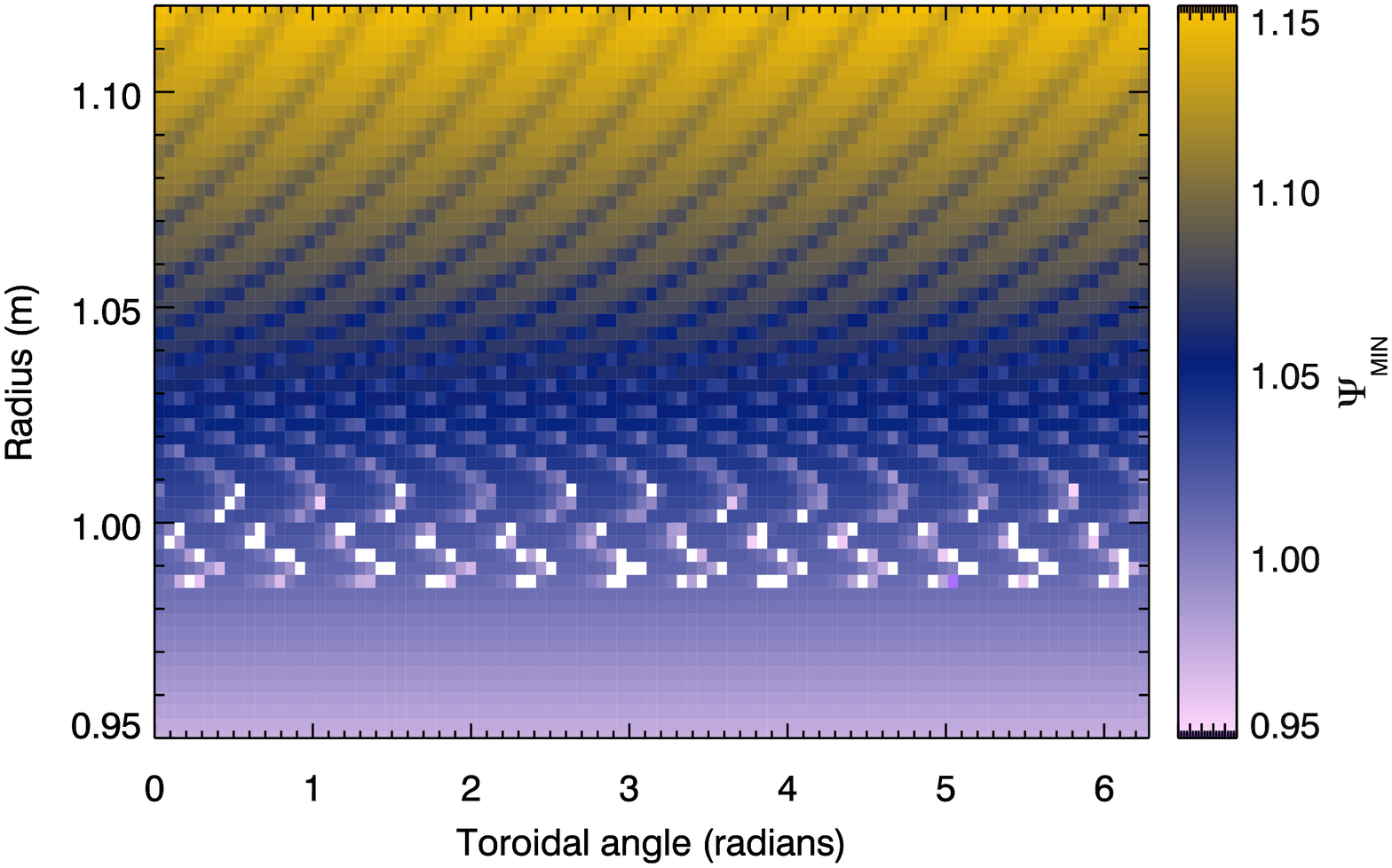}
        \caption{Modelling divertor strike point pattern for an ELM which is represented by 12 current carrying filaments. The plot shows the minimum $\Psi _{N}$ value a field line at a given position on the divertor reaches as a function of toroidal angle (x axis) and radius (y axis).}
        \label{fig:footprint_filaments_only}
\end{figure}

The presence of the current carrying ELM filaments gives rise to strike point splitting. Modelling of the strike point splitting using field line tracing through the ELM filament field shows clear splitting into 12 lobes, which is expected from the toroidal mode number of the filaments and shown in \fref{fig:footprint_filaments_only}.

The resonant component of the field from the ELM filaments is of the same magnitude as the applied n=3 RMP field. Therefore, the splitting from the RMP field will be affected by the presence of the ELM filaments. As the toroidal location of the ELM filaments varies from ELM to ELM, the phase between the n=12 ELM field and n=3 RMP field can vary from ELM to ELM. In the case where the ELM and RMP fields are in phase at the location of the IR camera, then the resonant components add and the splitting is amplified. In the case where the two fields are out of phase, this will result in the resonant components cancelling and the splitting not being present. The effect can be investigated by modelling the strike point splitting as the phase between the two fields is varied.

\begin{figure}[htp]
        \centering
        \includegraphics[width=0.4\textwidth]{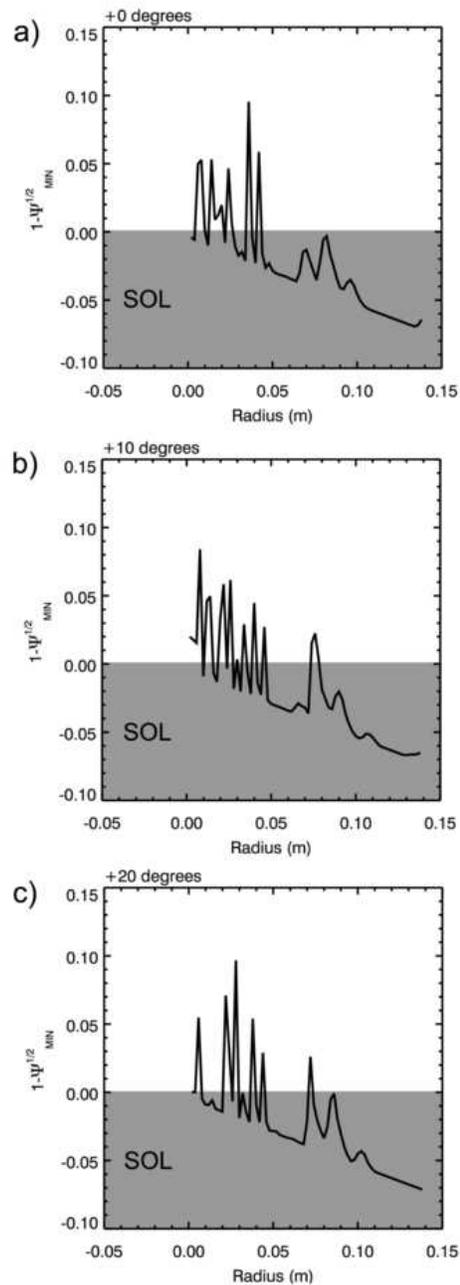}
        \caption{Modelled strike point splitting for three different phases of alignment between the ELM filament field and the RMP field.}
        \label{fig:filament_modelling}
\end{figure}

\Fref{fig:filament_modelling} shows the modelled field line excursion for three different phases of the ELM filament and RMP fields. The regions where the field line passes into the core, field line excursions above zero, are expected to receive a heat flux from the plasma. The first phase shown in \fref{fig:filament_modelling} a) suggests that only one lobe would be visible in the splitting pattern as there is a single region of field penetration into the core region between $\Delta R_{LCFS}$ 0 and 0.05m. The substructure present will be smeared by the instrument function of the IR camera (5 to 7.5 mm spatial resolution) which will merge the lobes together. If the phase between the coils is changed by 10 degrees, a secondary lobe is seen to form at a larger radius in the field line excursion profile, as shown in \fref{fig:filament_modelling} b). The secondary lobe can be seen to extend into the confined plasma region, and does not remain confined to the SOL, suggesting that a heat flux would be expected at this location on the divertor. Rotation of the phase of the two fields by a further 10 degrees suggests that the field lines in three of the lobes now extend out of the SOL and into the core plasma. The location of these lobes, $\Delta R_{LCFS}$ 0, 0.03-0.04 and 0.8 m, are consistent with the observed location of the measured splitting shown in \fref{fig:elm_modelled_cf}.

Further evidence for the splitting being generated by the interaction of the ELM filament and the RMP field is provided by the profiles of the individual ELMs (grey profiles) in \fref{fig:elm_cycle_profiles}. The splitting seen during the ELMs occurs at similar radial locations for all of the ELMs which exhibit splitting, but there is some variation in the location of the outer lobe which could suggest a locking with the RMP field at a different toroidal location. In addition, there is a weak dependence on the observation of the splitting with the ELM size which would be expected as an increased ELM size would produce a larger ELM current and a correspondingly larger field from the ELM filament which may promote locking. The data set is small, and further investigation is required to provide a conclusive result, however, the data shows that there is no splitting observed when the ELM energy is below 1 kJ and splitting is always seen when the ELM size is above 2.5 kJ. ELM energies within this range show a mixture of split and profiles without splitting.

\section{Conclusion}
\label{section:conclusion}

The divertor heat load generated by ELMs are a concern for ITER as a result of the cyclical nature of the power loading and the cracking of the divertor materials which can result. It is therefore necessary for ITER to have a form of ELM control, which can either suppress the ELMs completely, or mitigate them. ELM mitigation has been demonstrated on MAST using a set of in-vessel resonant magnetic perturbation (RMP) coils which act to increase the ELM frequency and thereby decrease the energy loss per ELM. ELM mitigation on MAST has been seen to lower the peak divertor heat flux \cite{thornton2013, kirk2013}, as required for ITER.

The application of the RMP to the plasma give rise to the formation of lobes at the X point of the plasma, and the interaction of these lobes with the divertor surfaces generate strike point splitting where the heat flux at the divertor develops side lobes. Strike point splitting is an important feature of RMP application, as it offers a means of studying the penetration of the RMP field into the plasma and assessing the level to which the plasma responds to the applied field. Increased levels of plasma response are expected to minimise the level of strike point splitting seen at the divertor, compared with that predicted by vacuum modelling. Measurements on MAST have been reported in both L and H-mode plasmas and show the evidence for strike point splitting. The splitting of the strike point is seen in both heat and particle fluxes to the divertor. The splitting is also seen during the ELM, where it is important to differentiate the splitting from the RMP and the striations formed as a result of the arrival of ELM filaments at the divertor. The ELM splitting is seen to vary during the ELM, with the largest splitting at the time of the peak heat flux. Comparison of the split profiles with vacuum modelling of the expected splitting are seen to be in moderate agreement in most cases. However, some of the predicted lobes are not seen in the measured profiles. Modelling of the strike point pattern which include screening resulting from the plasma rotation does show that the screening decreases the penetration of the RMPs into the plasma through an accompanied shortening of the lobes in toroidal extent. However, the screening is not seen to affect lobes containing field lines which remain confined to the scrape off layer (SOL). Therefore, the lack of the outer lobes predicted by the screened modelling in H-mode profiles must be due to the fact that there is an insufficient number of particles and heat reaching these outer lobes to deposited enough energy to the divertor for the lobe to be visible. 

The splitting during ELMs is seen to vary from ELM to ELM, with some ELMs showing splitting into three lobes, as expected from the toroidal mode number of the RMP, and others showing no splitting even at the same time during the ELM. A possible explanation for the variation is the relative alignment between the ELM mode, which have a toroidal mode number of n=12 in the example used here, and the RMP field which has a toroidal mode number of n=3. As the toroidal location at which the ELM is emitted is random, when the RMP and ELM filaments align then there is splitting and when they are out of alignment there is no splitting. Vacuum modelling utilising different phases between the ELM filaments and the RMP have been performed which support this hypothesis to explain the variation in the splitting.

\ack
This work was part funded by the RCUK Energy Programme [grant number EP/I501045] and the European Communities under the Contract of Association between EURATOM and CCFE. To obtain further information on the data and models underlying this paper please contact PublicationsManager@ccfe.ac.uk. The views and opinions expressed herein do not necessarily reflect those of the European Commission. The work of P. Cahyna was funded by the Grant Agency of the Czech Republic under grant P205/11/2341.

\section*{References}

\bibliographystyle{unsrt}
\bibliography{AJT_sfpw_paper_refs}

\end{document}